\documentclass{emulateapj}

\usepackage{ccaption,graphicx,amssymb,verbatim,times}

\newcommand\birx{IRX-$\beta$}

\slugcomment{}

\shorttitle{Dust Attenuation in UV-Selected Starbursts at High Redshift and their Local Counterparts}
\shortauthors{Overzier, R.A. et al.} 

\begin{document}



\title{Dust Attenuation in UV-selected Starbursts at High Redshift and
their Local Counterparts: Implications for the Cosmic Star Formation Rate Density}


\author{Roderik A. Overzier\altaffilmark{1}, 
Timothy M. Heckman\altaffilmark{2}, 
Jing Wang\altaffilmark{1},
Lee Armus\altaffilmark{3}, 
Veronique Buat\altaffilmark{4}, 
Justin Howell\altaffilmark{3}, 
Gerhardt Meurer\altaffilmark{5},  
Mark Seibert\altaffilmark{6}, 
Brian Siana\altaffilmark{7},
Antara Basu-Zych\altaffilmark{8}, 
St\'ephane Charlot\altaffilmark{9}, 
Thiago S. Gon\c calves\altaffilmark{7}, 
D. Christopher Martin\altaffilmark{7}, 
James D. Neill\altaffilmark{7}, 
R. Michael Rich\altaffilmark{10},
Samir Salim\altaffilmark{11}, 
David Schiminovich\altaffilmark{12}
}
\email{overzier@mpa-garching.mpg.de}

\altaffiltext{1}{Max-Planck-Institut for Astrophysics, D-85748 Garching, Germany.}
\altaffiltext{2}{Department of Physics and Astronomy, The Johns Hopkins University, 3400 North Charles Street, Baltimore, MD 21218.}
\altaffiltext{3}{Spitzer Science Center, Caltech, MS 220-6, Pasadena, CA 91125, USA.}
\altaffiltext{4}{Laboratoire d'Astrophysique de Marseille, OAMP, Université Aix-marseille, CNRS, 38 rue Frédéric Joliot-Curie, 13388 Marseille Cedex 13, France.}
\altaffiltext{5}{ICRAR/University of Western Australia, 35 Stirling Highway, Crawley, WA 6009, Australia.}
\altaffiltext{6}{Observatories of the Carnegie Institution of Washington, 813 Santa Barbara Street, Pasadena, California 91101, USA.}
\altaffiltext{7}{California Institute of Technology, MS 249-17, Pasadena, CA 91125, USA.}
\altaffiltext{8}{NASA Goddard Space Flight Center, Laboratory for X-ray Astrophysics, Greenbelt, MD 20771, USA.}
\altaffiltext{9}{PMC Univ Paris 06, UMR7095, Institut d'Astrophysique de Paris, F-75014 Paris, France.}
\altaffiltext{10}{Deptartment of Physics and Astronomy, Division of Astronomy and Astrophysics, University of California, Los Angeles, CA 90095-1562, USA.}
\altaffiltext{11}{National Optical Astronomical Observatories, 950 N. Cherry Avenue, Tucson, AZ 85719, USA.}
\altaffiltext{12}{Department of Astronomy, Columbia University, MC 2457, 550 West 120th Street, New York, NY 10027.}




\begin{abstract}
  We present a new analysis of the dust obscuration in starburst
  galaxies at low and high redshift.  This study is motivated by our
  unique sample of the most extreme UV-selected starburst galaxies in
  the nearby universe ($z<0.3$), found to be good analogs of
  high-redshift Lyman Break Galaxies (LBGs) in most of their physical
  properties.  We find that the dust properties of the Lyman Break
  Analogs (LBAs) are consistent with the relation derived previously
  by Meurer et al. (M99) that is commonly used to dust-correct star
  formation rate measurements at a very wide range of redshifts. We
  directly compare our results with high redshift samples (LBGs,
  ``$BzK$'', and sub-mm galaxies at $z\sim2-3$) having IR data
  either from Spitzer or Herschel. The attenuation in typical LBGs at
  $z\sim2-3$ and LBAs is very similar.  Because LBAs are much better
  analogs to LBGs compared to previous local
  star-forming samples, including M99, the practice of dust-correcting the SFRs of 
  high redshift galaxies based on the local calibration is now placed
  on a much more solid ground.  We illustrate the importance of this
  result by showing how the locally calibrated relation between UV measurements 
  and extinction is used to estimate the integrated, dust-corrected star
  formation rate density at $z\simeq2-6$.
   \end{abstract}



\section{Introduction}

The ultraviolet (UV) and far-infrared (far-IR) emission of
star-forming galaxies (SFGs) offers one of the most direct estimates
of their star formation rate (SFR). The fraction of light emerging in
the IR relative to that in the UV is modulated primarily by the amount
of dust seen by young stars
\citep[][]{meurer99,calzetti00,buat02,burgarella05,seibert05,treyer07},
with secondary effects due to, e.g., star formation history and
geometry
\citep[][]{charlot00,kong04,inoue06,boissier07,johnson07,salim07,panuzzo07,cortese08,boselli10}. For
a relatively young galaxy forming stars at a stable rate, the SFR is
proportional to the UV luminosity \citep{leitherer95}, while the
degree of reddening of the UV continuum is related to the amount of
dust. This reddening can be expressed in terms of the UV slope,
$\beta$ with $f_\lambda\propto\lambda^\beta$, and the total
attenuation is given by the ratio $IRX{\equiv}L_{IR}/L_{UV}$
\citep[][M99]{meurer99}. The ``$IRX-\beta$'' relation of M99 is based
on a large sample of local starburst galaxies, and has proven to be an
extremely useful tool for estimating total SFRs from UV-only data. For
the relation to work, it is implied that the dust must be near the UV
sources and have some kind of shell- or screen-like geometry
\citep{gordon00}.

The IRX relation is particularly important at high
redshift. SFR estimates based on IR, X-ray, or radio data are typically
available only for the brightest objects
\citep{pope06,siana08,siana09,reddy06,reddy10}, or statistically
through stacks
\citep[][]{seibert02,carilli08,reddy10,magdis10a,magdis10b,ho10,kurczynski10,rigopoulou10}. Surveys
in the rest-frame UV can simultaneously give $L_{FUV}$ and
$\beta$ and hence the dust-corrected SFR based on the local IRX
relation. Using this technique, \citet[][B09]{bouwens09} estimated the
$z\sim3-6$ SFR density from large samples of LBGs. They found that most of the energy output at
$z=2.5-4$ ($z=4-6$) occurs in the IR (UV), and that the SFR density can be largely recovered 
by dust-correcting the UV measurements thus demonstrating the
importance of UV surveys. However, because local calibrations of the
IRX relation are based on galaxies that are very different from
typical UV-selected starburst galaxies at $z\gtrsim3$, it is not clear
whether they are valid at high redshift. In this Letter, we
re-investigate the \birx\ relation for the first time using a unique
population of nearby ($z<0.3$) starbursts that has been 
shown to be similar to LBGs in most of their basic physical
properties. These ``Lyman Break Analogs'' (LBAs) are similar in mass,
age, size, metallicity, optical extinction, and SFR
\citep{heckman05,hoopes07,basu07}, have similar compact and
clumpy morphologies \citep{overzier08,overzier10}, similar kinematics
\citep{basu09,goncalves10}, and a similar feedback-dominated interstellar medium \citep[][O09]{overzier09}.
The structure of this Letter is as follows. We first present our data and
measurements (\S2).  We compare LBAs with low and high redshift SFGs
having good IR data from Spitzer or Herschel (\S3), and we discuss the
possible implications of our results on the SFR density at high redshift (\S4).

\begin{figure*}[t]
\begin{center}
\includegraphics[width=0.8\textwidth]{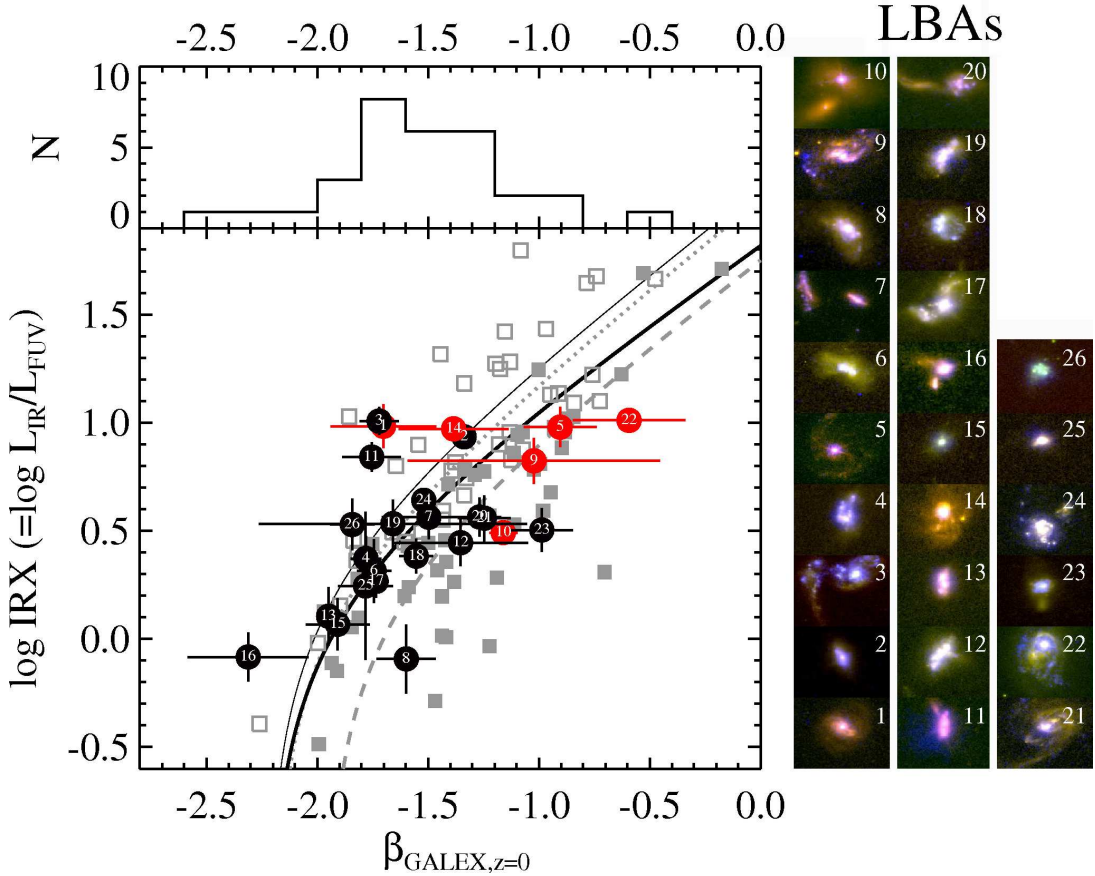}
\end{center}
\caption{\label{fig:irx}\textit{Main panel:} The relation between the UV
  slope and ``IRX'' for LBAs (\textit{large symbols}). LBAs referred to as ``Dominant Compact Objects'' (DCOs, O09)
  are marked in red, and HST morphologies from O09 are shown on the right.  Open and filled squares indicate M99
  galaxies re-measured using GALEX data (\S2.2) using a small
  aperture similar to the original IUE measurements (open squares), and
  using an aperture enclosing the entire galaxy (filled squares). Fits
  to the data are indicated (\textit{thick black line}: IRX$_{LBA}$; \textit{grey dotted line}: 
  IRX$_{M99,inner}$; \textit{grey dashed line}: IRX$_{M99,total}$). The original relation from M99, IRX$_{M99,0}$, is shown for comparison (\textit{thin black
  line}). \textit{Top panel:} Distribution in $\beta_{z=0}$ for LBAs.}
\end{figure*}

\begin{figure*}[t]
\begin{center}
\includegraphics[width=0.7\textwidth]{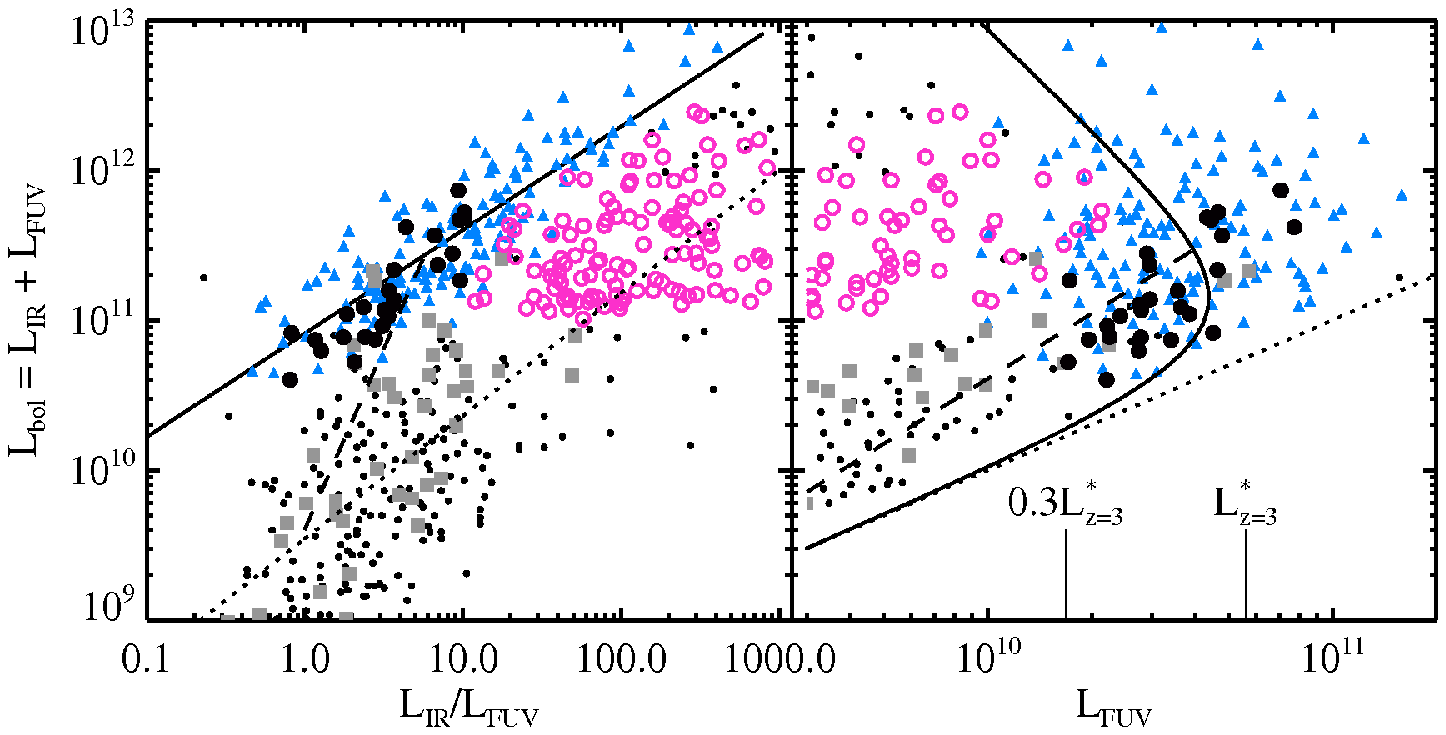}
\end{center}
\caption{\label{fig:lbol}$L_{bol}$ versus the attenuation
  ($L_{IR}/L_{FUV}$, left panel) and $L_{FUV}$ (right panel) for LBAs
  (large circles), $z\sim2$ LBGs (blue triangles, from R10), M99 (grey squares,
  with $R<R_{max}$), LIRGs (open magenta circles, \citet{howell10}), and
  ordinary SFGs (small circles, see \citet{huang09}). The local
  samples (M99, SFGs and LIRGs) lie along a broad relation between
  $L_{bol}$ and attenuation (the result of a simple linear fit to guide the eye is
  indicated by the dotted line). The most extreme UV-selected samples
  at low (LBAs) and high redshift (LBGs) also lie along such a
  relation, but one that is offset by 1--2 dex towards lower
  attenuations with respect to the former. The mean relation at
  $z\sim2$ from R10 is indicated (solid: BM/BX sample; dashed:
  extrapolation to intrinsically faint objects). The
  fraction of $L_{bol}$ emitted in the FUV is much
  higher for LBGs and LBAs compared to typical local SFGs (right panel).}
\end{figure*}

\section{Data and Methods}

\subsection{Lyman Break Analogs UV and IR data}

The sample of 31 LBAs from O09 forms the basis of this paper. FUV
  luminosities, $L_{FUV}\equiv{\lambda}L_\lambda$ with
  $\lambda=1600$\AA, were calculated from our F150LP
  ($\lambda_c$$\approx$1614\AA) data taken with the \textit{Hubble
    Space Telescope} (HST) Advanced Camera for Surveys
  \citep[see][O09, for details]{overzier08}. The slope of the UV
  continuum, $\beta_{GALEX}$, was
  calculated\footnote{$\beta_{GALEX}=\frac{0.4(m_{FUV}-m_{NUV})}{\mathrm{log}_{10}(\lambda_{FUV}/\lambda_{NUV})}-2.0$}
  from Galaxy Evolution Explorer (GALEX) General Release 6 data in the
  FUV ($\lambda_c$$\approx$1530\AA) and NUV
  ($\lambda_c$$\approx$2315\AA) having total exposure times of
  $\sim$100--13,000 s. All LBAs are unresolved in the GALEX images,
  and we measured total magnitudes in a 9\arcsec\ radius aperture.  We
  perform a small $K$-correction to obtain $L_{FUV}$ and
  $\beta_{GALEX}$ at $z=0$ using a set of starburst templates. In this
  paper we work exclusively with the photometric measure of $\beta$
  (as opposed to the ``true'' $\beta$ obtained from a power-law fit to
  a UV spectrum), that is most appropriate for comparing with samples for
which spectra are usually not available in large numbers.

We use IR data obtained with the Infrared Array Camera (IRAC), the
Multi-band Imaging Photometer (MIPS), and the InfraRed Spectrograph
(IRS) on the {\it Spitzer Space Telescope} (O09, Armus et al. in
prep.). At 24$\mu$m and shorter wavelengths, point source
photometry was performed on the post-Basic Calibration Data
(``pBCD''). At 70 and 160$\mu$m, flux densities were obtained by
aperture photometry on the filtered pBCD images. Upper limits were
determined from the standard deviation images. The IRS spectra
covering observed wavelengths in the range 5--35$\mu$m were normalized
to the MIPS 24$\mu$m data.
We estimate IR luminosities, $L_{IR}$ (3-1000$\mu$m), by fitting our
data with the model library of \citet{siebenmorgen07} (SK07). This
method allows us to fit models covering a wide range of physical
parameters, thus obtaining a good sense of the range of models allowed
by our data within the measurement errors.  We simulate our
IRAC+MIPS+IRS data set (including measurement errors and upper limits)
and calculate $L_{IR}$ and its error by taking the median and standard
deviation of all the templates that best fit the data in the monte
carlo simulation\footnote{While our results obtained using SK07 are
  consistent with those obtained from black-body fitting, we note that
  the SK07 library generally performs better for LBAs compared to the
  more commonly used empirical library from \citet{chary01}. The
  latter often failed to simultaneously fit the mid-IR and the far-IR
  dust emission, presumably due to the fact that the LBAs show a
  greater spectral variation than typical IR-selected starbursts in
  the local universe.}. Four objects lacking a sufficient number of IR 
data points were removed from the sample, as well as one object with
an (obscured) active nucleus identified in recent optical
spectroscopy. The UV and IR measurements for the 26 remaining objects
are given in Table \ref{tab:photprops}.

\subsection{Redetermination of the Meurer et al. (1999) relation}
\label{sec:jing} 

In order to be able to compare the results from different samples at
low and high redshift in a consistent manner, we have re-measured the
original M99 \birx\ relation using GALEX data and an updated estimate
of $L_{IR}$. 47 galaxies of the M99 sample are covered by
GALEX. Details on the photometry will be provided in a forthcoming
paper. In brief, we follow the procedures in \citet{wang10} carefully
masking neighbours and image artefacts, perform background
subtraction, and photometry on PSF-matched images using elliptical
apertures out to a maximum radius, $r_{max}$, of
$2.5-5{\times}r_{kron}$. Because $r_{max}$ is in most cases much
larger than the International UV Explorer (IUE) aperture of
$10\arcsec\times20\arcsec$ used by M99, we also measure the inner UV
colors and fluxes within a maximum radius of 10\arcsec\ for comparison
with earlier work. We compute $L_{IR}$ using the Infared Astronomical
Satellite (IRAS) fluxes at 12, 25, 60, and 100$\mu$m
\citep{sanders96}.

\section{Results}

\subsection{The \birx\ Relation of LBAs versus M99}

In Fig. \ref{fig:irx} we show the \birx\ diagram for LBAs. Panels show
the corresponding HST morphologies from O09, and the
distribution of UV slopes (top panel). We also show the values
measured in \S\ref{sec:jing} for the M99 galaxies using an aperture
radius of $10\arcsec$ most similar to M99, and within $R_{max}$
enclosing the entire source. We follow M99 and write IRX in terms of
$A_{FUV}$, the attenuation at 1600\AA, and the bolometric corrections
to the total light emitted by stars (BC$_{FUV,*}$) and dust (BC$_{dust}$):
\begin{equation}
\mathrm{log}_{10}(IRX)=\mathrm{log}_{10}(10^{0.4A_{FUV}}-1)+\mathrm{log}_{10}\frac{\mathrm{BC}_{FUV,*}}{\mathrm{BC}_{dust}}, 
\end{equation}
with BC$_{FUV,*}\approx1.68$ \citep[M99,][]{seibert05}, and
BC$_{dust}\approx1$ for our estimate of $L_{IR}$. We then derive the
best-fit relation by performing a linear fit of the form
$A_{FUV}=C_0+C_1\beta$ to the data in Fig. \ref{fig:irx}. Similar to
M99 we exclude galaxies having radii of ${>}2\arcmin$. We find
$A_{FUV}=4.54+2.07\beta\pm0.4$ (IRX$_{M99,inner}$, dotted line), and
$A_{FUV}=3.85+1.96\beta\pm0.4$ (IRX$_{M99,total}$, dashed line). Our
redetermination of IRX$_{M99,inner}$ is very similar to the original
M99 relation (IRX$_{M99,0}$, thin solid line).  However, using the new
larger aperture we find $\sim2\times$ more flux in the FUV for nearly
all sources, and hence a lower IRX. The smaller IUE aperture used by
M99 may thus have missed about half the light. On the other hand, in
both cases we use a single integrated IR luminosity of unknown spatial
distribution. For a subset of 12 galaxies from M99, we were able to
compare the IRAS measurements with more recent determinations of $L_{IR}$ based on Spitzer photometry encompassing the
entire source \citep{engelbracht08}. The results are in good
agreement. If a significant fraction of the IR is due to
heating by stars further out, using IRX$_{M99,0}$ rather than
IRX$_{M99,total}$ would lead to an overestimate of the total
attenuation. Alternatively, IRX$_{M99,total}$ could be affected by the
(redder) UV light from somewhat older stellar populations on the
outskirts of the galaxies. In this case, the interpretation of
IRX$_{M99,total}$ would be less straightforward as it
requires knowledge of the star formation history
\citep[see][]{kong04}.

How do LBAs compare to these IRX relations? A fit to LBAs gives
$A_{FUV}=4.01+1.81\beta$ (implied FUV attenuations of 0--3 mag).  The
dispersion found in IRX$_{LBA}$ is $\sim$0.6. As shown in
Fig. \ref{fig:irx}, the LBAs are thus in much better agreement, on
average, with M99 (inner) than with M99 (total), but we note that this
only applies to the range of $-2.5\lesssim\beta\lesssim-1$ probed by
these samples. We conclude that the dust properties of LBAs are most similar to 
those found only in the inner starburst cores of the M99 galaxies.

\subsection{The Attenuation of Starbursts at Low and High Redshift}

With good estimates of $L_{IR}$ and $L_{FUV}$ and thus the bolometric
luminosity ($L_{bol}{\equiv}L_{FUV}+L_{IR}$), we can compare LBAs to
other local and high redshift SFGs. This is shown in
Fig. \ref{fig:lbol}. In the left panel we show $L_{bol}$ versus the
attenuation ($L_{IR}/L_{FUV}$) for LBAs, ordinary SFGs, M99
starbursts, and Luminous IR Galaxies (LIRGs) all at low redshift, and
``BM/BX'' LBGs at $z\sim2$ from R10.  Typical SFGs in the local universe lie along a broad sequence
in which the attenuation roughly follows $L_{bol}$ (indicated by the
dotted line). LBAs and LBGs follow a similar relation, but one
that is offset by 1--2 dex toward lower attenuations at the same
$L_{bol}$ (solid line shows the best-fit relation from R10). Conversely, at fixed $L_{bol}$ LBAs and LBGs reach much higher FUV
luminosities than typical local star-forming galaxies
(i.e. $L_{FUV}\gtrsim0.3L^*_{z=3}$, right panel of
Fig. \ref{fig:lbol}).

\begin{figure*}[t]
\begin{center}
\includegraphics[width=0.7\textwidth]{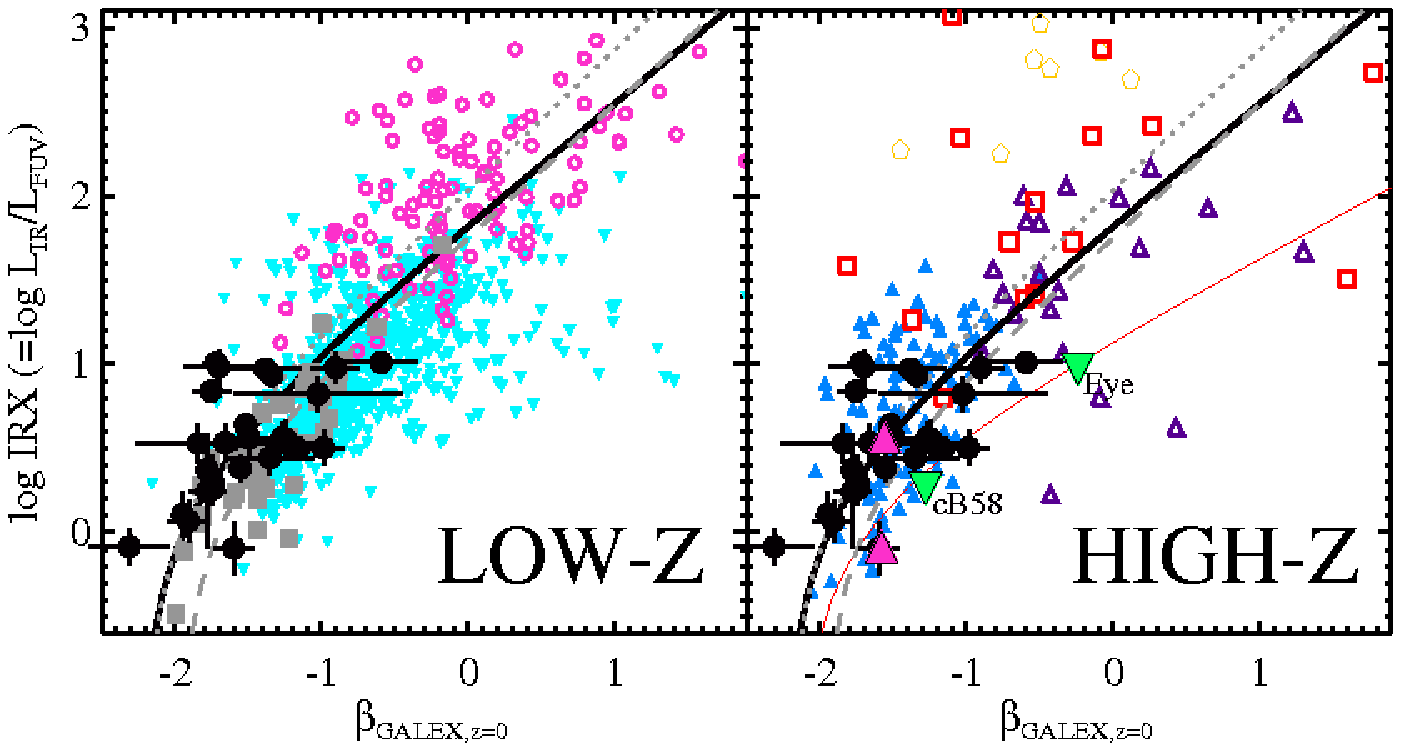}
\end{center}
\caption{\label{fig:irx2}IRX-$\beta$ at low
  (left) and high redshift (right). LBAs are indicated in both panels (black points). {\it Left panel:} M99 starbursts with
  $R<R_{max}$ (grey squares), SFGs (cyan upside-down triangles;
  \citet{buat10}), and (U)LIRGs (open magenta circles;
  \citet{howell10}. {\it Right panel:} LBGs (filled blue triangles),
  BzK galaxies (open purple triangles), distant red galaxies (open red
  squares) and SMGs (open orange pentagons) at $z\sim2-3$
  \citep[][R10]{reddy06}, lensed LBGs at $z\sim3$ (large green
  triangles; \citet{siana08,siana09}, and statistical detections of
  LBGs at $z\sim3$ detected at 1.1mm \citep[lower magenta
  triangle,][]{magdis10a} or at 160$\mu$m with Herschel \citep[upper
  magenta triangle,][]{magdis10b}. Curves show IRX$_{M99,inner}$
  (dotted), IRX$_{M99,total}$ (dashed) and IRX$_{LBA}$ (thick black)
  from Fig. \ref{fig:irx} and \S3.1. The IRX relation assuming a SMC
  extinction curve is indicated by the thin red line.}
\end{figure*}

Because LBAs appear so similar compared to $z\sim2$ LBGs in terms of
their $L_{bol}$ and attenuation (Fig. \ref{fig:lbol}), it is
interesting to see whether the IRX relation at low and high redshift
is similar as well. In Fig. \ref{fig:irx2} we show a compilation of
low and high redshift data for which accurate measurements are
available. We first compare LBAs with the sample of $z<0.2$ SFGs
detected with GALEX and Herschel from \citet{buat10}, and with LIRGs
from \citet{howell10}, shown left. Aperture effects either in the UV
or the IR are most likely much less of a problem compared to the M99
sample, due to the higher average redshifts of the two samples
compared to that of M99. Therefore we will use IRX$_{M99,total}$ as
our reference (dashed line). LIRGs tend to lie above this relation,
while the galaxies from \citet{buat10} lie below it (on average). For
these samples, we would therefore tend to, respectively, under- and
overestimate the attenuation when using IRX$_{M99,total}$, while
IRX$_{M99,inner}$ performs somewhat better for LIRGs \citep[see][for
detailed discussion]{buat10,howell10}.

In the right panel of Fig. \ref{fig:irx2} we present an overview of
the situation at $z>2$, showing LBGs, $BzK$ and sub-mm galaxies (SMGs)
at $z=2-3$ \citep[][R10]{reddy06}, lensed LBGs \citep[``cB58'' and
the cosmic ``Eye'';][]{siana08,siana09}, and stacked LBGs at $z\sim3$
\citep{magdis10a,magdis10b}. LBAs and $z\sim2$ LBGs occupy a very
similar region in this IRX-$\beta$ diagram, confirming the similarity
between the two samples. We should note that the IRX estimate of R10
is not based on a direct measurement of the IR emission (it is based
on a combination of $L_{FUV}$, $L_{H\alpha}$ and $L_8$), but
statistical detections in the X-rays appear consistent with the
extrapolated estimates of $L_{IR}$ \citep[][R10]{seibert02}. We can also compare our
results with the statistical detection of $z\sim3$ LBGs in stacks at
100$\mu$m and 160$\mu$m from recent Herschel observations.  This
result is indicated by the upper magenta triangle, which falls right
in the middle of the distribution of LBAs and $z\sim2$ LBGs. This
suggests that even for the most IR-luminous LBGs (median $L_{IR}$ of
$1.6\times10^{12}L_\odot$, i.e. ULIRGs) the attenuation is exactly as expected
based on the locally determined IRX (inner) relation. 

\begin{figure*}[t]
\begin{center}
\includegraphics[width=0.7\textwidth]{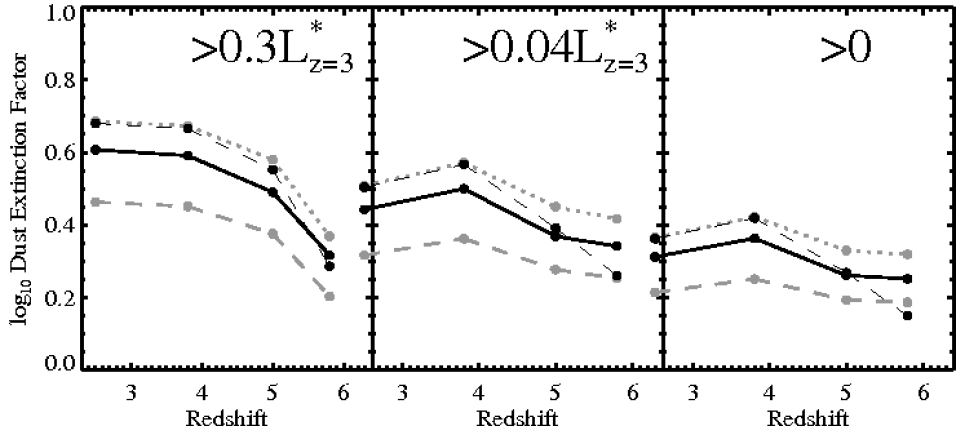}
\end{center}
\caption{\label{fig:b09}The multiplication factors needed to
  dust-correct the SFR density at $z\sim3-6$ based on the UV slopes
  and luminosity functions (LF) of $U,B,V,i$ ``dropout'' galaxies from B09,
  and assuming that the local attenuation corrections are valid at
  high redshift. Panels shows the total dust corrections implied when
  integrating the LFs over ($L/L^*_{z=3}$)$>$0.3, 0.04, and 0. The curves show the results when using
  IRX$_{LBA}$ (black solid line), IRX$_{M99,inner}$ (grey dotted line) and IRX$_{M99,total}$
  (grey dashed line) from \S3.1.  The thin dashed curve shows the results of B09 using
  the original M99 relation (IRX$_{M99,0}$).}
\end{figure*}

Direct measurements of the mid- or far-IR emission from LBGs are
available in only a handful of cases. In Fig. \ref{fig:irx2} we
indicate the results found for two bright, lensed LBGs at $z\sim3$
from \citet[][]{siana08,siana09}. These objects have direct detections
at (observed) 24 and 70$\mu$m (as well as 850$\mu$m and 1.2 mm in the
case of cB58), allowing a good estimate of $L_{IR}$.  It has been
noted that both objects appear to lie substantially below
IRX$_{M99,0}$, perhaps requiring a modified extinction curve or an
extremely young stellar population \citep[see][and the discussion on
``young galaxies'' by R10]{baker01,siana09}. An extinction curve more
similar to that of the Small Magellanic Cloud (SMC), i.e., extinction
that occurs in a more uniform rather than a patchy foreground dust
screen would require less emission by dust for the same stellar
population age, and hence lower IRX (see red curve in
Fig. \ref{fig:irx2}). 
cB58 lies on the lower envelope formed by LBAs and $z\sim2$ galaxies,
and the observed scatter is larger than the difference expected
between the standard Calzetti-type and SMC-type extinction curves for small $\beta$. 
The cosmic Eye, however, has a much larger slope ($\beta\approx-0.3$) and lies well below the
starburst IRX relations. Previous studies have shown that offsets from
IRX are commonly due to more complicated star formation histories
\citep[e.g.][]{kong04}. This is a likely explanation, at least for the
local SFGs (left panel), BzK galaxies \citep[right panel, see
also][]{nordon10}, and LBAs (see objects marked ``DCOs'' in
Fig. \ref{fig:irx} and O09) having relatively low IRX ratios. For the
cosmic Eye, however, a modified extinction curve may indeed be
required in order to simultaneously explain its relatively simple
stellar population and its low IRX ratio \citep{siana09}.

\section{Discussion and conclusions}

The first goal of this paper was to investigate whether LBAs are
consistent with the M99 relation, and we have shown that this is
indeed the case. This is an important result: Because LBAs are much
better analogs to LBGs compared to previous local samples, including
M99, the applicability of the local dust correction at high redshift
now appears to be on much more solid ground. Confirmation of
this result is given by the fact that LBAs are also very similar to
typical LBGs in terms of their dust properties (Figs. \ref{fig:lbol}
and \ref{fig:irx2}), although some caution must be taken as true 
$L_{IR}$ measurements are still sparse at high redshift.

We can show the implications of these results on the determination of
the SFR density at $z\sim3-6$. B09 used large dropout samples from the
Hubble Ultra Deep Field to quantify the relation between $\beta_{UV}$
and $M_{UV}$, and computed the total dust corrections to the SFR
density by integrating the UV luminosity functions at each redshift
\citep{bouwens07,reddy09} and assuming that IRX$_{M99,0}$ is
valid. The method and data that we use here is similar to that of B09
(Fig.~8 and \S5.4 of that work), but instead of using IRX$_{M99,0}$ we
show the results for our new IRX relations derived in \S3.1. The
results are shown in Fig. \ref{fig:b09}, which gives the total
correction factor that is needed to dust-correct a measurement of the
UV luminosity (or SFR) density at $z\sim3-6$. Not surprisingly,
IRX$_{LBA}$ (thick black lines) and IRX$_{M99,inner}$ (dotted lines)
give quite similar results to the relation used by B09 (thin dashed
lines). Note that if we assume that IRX$_{M99,total}$ measured within
the larger aperture provides a better estimate of the total
attenuation, the total dust-correction would have to be lowered by
$\sim0.2$ dex at all redshifts and for all luminosities (grey dashed
lines). We conclude that current determinations of the cosmic SFR
density based on dust-corrected UV data from Lyman break galaxy
surveys appear to be on solid ground, but, as noted by B09, this is
mainly due to the large contribution from very faint, blue sources at
high redshift. 

Direct measurements of $L_{IR}$ for relatively red ($\beta\gtrsim0$)
galaxies at $z\gtrsim2-3$ are needed to constrain the IRX relation in
this regime, while the most obscured sources typically do not follow
the IRX relations at all (e.g. ULIRGs at low and SMGs at high
redshift). This illustrates the importance of surveys in the IR
currently being performed by Herschel. In forthcoming papers we will
present a more detailed analysis of the dust properties of LBAs and
LBGs based on IRS spectra as well as new far-IR data from Herschel.

\acknowledgments{We thank Guinevere Kauffmann, Rychard Bouwens, Luca Cortese and Ranga-Ram Chary for providing useful comments.}

{}

\begin{deluxetable*}{lcccccccccc}
\tabletypesize{\tiny}
\tablecolumns{11}
\tablewidth{0pc}
\tablecaption{\label{tab:photprops}Lyman Break Analogs: UV and IR Measurements.}
\tablehead{
\multicolumn{1}{c}{ID$^a$} &
\multicolumn{1}{c}{Name} & 
\multicolumn{1}{c}{$z$} & 
\multicolumn{1}{c}{$m_{FUV}$} &  
\multicolumn{1}{c}{$m_{NUV}$} &  
\multicolumn{1}{c}{$\beta_{obs}$}  & 
\multicolumn{1}{c}{$\beta_{z=0}$}  & 
\multicolumn{1}{c}{log $L_{FUV}$}  & 
\multicolumn{1}{c}{log $L_{FIR}$$^b$} & 
\multicolumn{1}{c}{log $L_{IR}$$^c$} & 
\multicolumn{1}{c}{log IRX}\\ 
& (SDSSJ...) & & (mag) & (mag) &  &&  ($L_\odot$)  & ($L_\odot$)  & ($L_\odot$) & }
\startdata
       1$^\dagger$ & 005439.79+155446.9 & 0.236 & 20.47 & 20.23 & $-1.44\pm0.24$ & $-1.69$ & $10.24\pm0.02$   &  $11.03\pm0.10$ &$11.22\pm0.10$ & $0.98\pm0.10$\\
    2 & 005527.46-002148.7 & 0.167 & 19.08 & 18.71 & $-1.15\pm0.05$ & $-1.33$ & $10.46\pm0.01$                      &  $11.04\pm0.08$  &  $11.40\pm0.05$ & $0.93\pm0.05$\\
    3 & 015028.40+130858.3 & 0.147 & 18.48 & 18.33 & $-1.64\pm0.09$ & $-1.71$ & $10.64\pm0.01$                      & $11.40\pm0.08$   & $11.64\pm0.06$ & $1.01\pm0.07$\\
    4 & 020356.91-080758.5 & 0.189 & 19.15 & 19.00 & $-1.65\pm0.07$ & $-1.77$ & $10.56\pm0.01$                      & $10.68\pm0.10$   & $10.93\pm0.10$ & $0.37\pm0.10$\\
    5$^\dagger$ & 021348.53+125951.4 & 0.219 & 19.60 & 18.91 & $-0.46\pm0.16$ & $-0.89$ & $10.65\pm0.02$      &  $11.38\pm0.08$  & $11.63\pm0.09$ & $0.98\pm0.09$\\
    6 & 032845.99+011150.8 & 0.142 & 19.19 & 19.05 & $-1.68\pm0.08$ & $-1.73$ & $10.23\pm0.01$                     &  $10.31\pm0.17$   & $10.55\pm0.15$ & $0.31\pm0.15$\\
    7 & 035733.99-053719.7 & 0.204 & 19.65 & 19.33 & $-1.27\pm0.21$ & $-1.49$ & $10.47\pm0.02$                      &  $10.81\pm0.10$  & $11.03\pm0.10$ & $0.56\pm0.10$\\
    8 & 040208.86-050642.0 & 0.139 & 18.89 & 18.68 & $-1.51\pm0.13$ & $-1.59$ & $10.34\pm0.02$                      &  $10.03\pm0.20$  & $10.25\pm0.16$ & $-0.09\pm0.16$\\
    9$^\dagger$ & 080232.34+391552.6 & 0.267 & 20.02 & 19.34 & $-0.45\pm0.57$ & $-1.01$ & $10.68\pm0.06$     &  $11.23\pm0.11$  & $11.50\pm0.09$ & $0.82\pm0.11$\\
   10$^\dagger$ & 080844.26+394852.3 & 0.091 & 17.98 & 17.59 & $-1.10\pm0.06$ & $-1.15$ & $10.35\pm0.01$    &  $10.47\pm0.07$  &   $10.84\pm0.03$ & $0.49\pm0.03$\\
   11 & 082001.72+505039.1 & 0.217 & 19.85 & 19.65 & $-1.55\pm0.13$ & $-1.74$ & $10.47\pm0.01$                   &  $10.98\pm0.10$  & $11.31\pm0.07$ & $0.84\pm0.07$\\
   12 & 082550.95+411710.2 & 0.156 & 19.58 & 19.23 & $-1.19\pm0.31$ & $-1.34$ & $10.29\pm0.02$                   & $10.53\pm0.09$   & $10.74\pm0.11$ & $0.45\pm0.11$\\
   13 & 083803.72+445900.2 & 0.143 & 18.92 & 18.89 & $-1.92\pm0.04$ & $-1.94$ & $10.44\pm0.00$                   & $10.32\pm0.11$   & $10.54\pm0.13$ & $0.10\pm0.13$\\
   14$^\dagger$ & 092159.38+450912.3 & 0.235 & 19.19 & 18.76 & $-1.02\pm0.25$ & $-1.37$ & $10.85\pm0.02$   &  $11.57\pm0.06$  &  $11.82\pm0.05$ & $0.97\pm0.05$\\
   15 & 092336.45+544839.2 & 0.222 & 19.72 & 19.61 & $-1.75\pm0.14$ & $-1.90$ & $10.53\pm0.01$                   &   $10.27\pm0.18$ & $10.60\pm0.12$ & $0.07\pm0.12$\\
   16 & 092600.40+442736.1 & 0.181 & 18.69 & 18.83 & $-2.32\pm0.27$ & $-2.30$ & $10.65\pm0.02$                   &  $10.19\pm0.17$  & $10.57\pm0.11$ & $-0.08\pm0.11$\\
   17 & 093813.49+542825.0 & 0.102 & 17.83 & 17.72 & $-1.73\pm0.05$ & $-1.72$ & $10.59\pm0.00$                  &   $10.51\pm0.09$  & $10.85\pm0.08$ & $0.27\pm0.08$\\
   18 & 102613.97+484458.9 & 0.160 & 19.20 & 18.95 & $-1.43\pm0.08$ & $-1.54$ & $10.35\pm0.01$                  &   $10.50\pm0.08$  & $10.74\pm0.08$ & $0.38\pm0.08$\\
   19 & 124819.74+662142.6 & 0.260 & 19.98 & 19.70 & $-1.35\pm0.60$ & $-1.65$ & $10.55\pm0.06$                  &   $10.81\pm0.11$  & $11.08\pm0.10$ & $0.53\pm0.11$\\
   20 & 135355.90+664800.5 & 0.198 & 18.99 & 18.54 & $-0.98\pm0.10$ & $-1.26$ & $10.67\pm0.02$                  &   $11.05\pm0.08$ & $11.23\pm0.09$ & $0.56\pm0.09$\\
   21 & 143417.15+020742.3 & 0.180 & 19.50 & 19.07 & $-1.01\pm0.12$ & $-1.24$ & $10.44\pm0.01$                  &   $10.72\pm0.11$ & $11.00\pm0.11$ & $0.56\pm0.11$\\
   22$^\dagger$ & 210358.74-072802.4 & 0.137 & 18.49 & 17.76 & $-0.35\pm0.25$ & $-0.58 $ & $10.67\pm0.01$  &  $11.36\pm0.06$ & $11.68\pm0.03$ & $1.01\pm0.04$\\
   23 & 214500.25+011157.5 & 0.204 & 20.00 & 19.40 & $-0.63\pm0.14$ & $-0.98 $ & $10.45\pm0.01$                  &  $10.67\pm0.13$ & $10.95\pm0.10$ & $0.50\pm0.10$\\
   24 & 231812.99-004126.1 & 0.252 & 19.23 & 18.87 & $-1.18\pm0.04$ & $-1.51 $ & $10.89\pm0.00$                  &   $11.17\pm0.09$ & $11.53\pm0.06$ & $0.64\pm0.06$\\
   25 & 232539.22+004507.2 & 0.277 & 20.47 & 20.25 & $-1.49\pm0.12$ & $-1.77 $ & $10.44\pm0.01$                  &  $10.31\pm0.48$  & $10.69\pm0.34$ & $0.25\pm0.34$\\
   26 & 235347.68+005402.0 & 0.223 & 20.07 & 19.92 & $-1.66\pm0.10$ & $-1.83$ & $10.38\pm0.01$                   & $10.62\pm0.18$  & $10.91\pm0.12$ & $0.53\pm0.12$
\enddata
\tablenotetext{a}{ID refers to the labels shown in Figure 1.}
\tablenotetext{b}{Far-IR luminosity integrated over the range 40-120$\mu$m.}
\tablenotetext{c}{IR luminosity integrated over the range 3-1000$\mu$m.}
\tablenotetext{$\dagger$}{Indicates objects referred to as Dominant Compact Objects in O09.}
\end{deluxetable*}

\end{document}